\documentclass[twocolumn,amsmath,amssymb,aps,prl,longbibliography, floatfix]{revtex4-2}
\usepackage{physics}
\usepackage{units}
\usepackage{amsmath}
\usepackage{url}
\usepackage{natbib}
\usepackage{textcase}
\usepackage{amssymb}
\usepackage{graphicx}
\usepackage{bm} 
\usepackage{soul,color}
\setstcolor{red}
\usepackage{times}
\usepackage{float}
\usepackage{multirow,microtype,color,relsize,ulem}
\usepackage{subfigure}
\usepackage[breaklinks=true]{hyperref}
\usepackage[utf8]{inputenc}
\usepackage[english]{babel}
\hypersetup{colorlinks=true,linkcolor=blue,citecolor=blue}
\hypersetup{linktocpage}
\usepackage{CJK}
\usepackage{breakcites}
\usepackage{float}
\usepackage{siunitx}
\usepackage[dvipsnames]{xcolor}
\usepackage{comment}

\def\Vsd{\tau_{\rm q}}

\definecolor{zycolor}{RGB}{60,179,113}

\definecolor{vspcolor}{RGB}{250,10,103}

\def\Vsd{V_{\rm sd}}

\def\Eh{E_{\rm H}}
\def\Eb{E_{\rm B}}
\def\Ez{E_{\rm Z}}
\def\Ei{E_{\rm I}}
\def\Eso{E_{\rm SO}}
\def\E{E}
\def\Vl{V_{\rm L}}
\def\Vr{V_{\rm R}}
\def\Mul{\mu_{\rm L}}
\def\Mur{\mu_{\rm R}}
\def\Guu{G_{\rm \uparrow\uparrow}}
\def\Gud{G_{\rm \downarrow\uparrow}}
\def\Gdu{G_{\rm \uparrow\downarrow}}
\def\Gdd{G_{\rm \downarrow\downarrow}}
\def\Go{e^2/h}

\def\tmaj{t_{\rm maj}}
\def\tmin{t_{\rm min}}

\def\Bx{B_{\rm x}}
\def\Bso{\vec{B}_{\rm SO}}

\begin{document}

\title{Spin-helical detection in a semiconductor quantum device with ferromagnetic contacts}
\author{Zedong Yang}
\author{Paul A. Crowell}
\author{Vlad S. Pribiag}
    \email{vpribiag@umn.edu}
\affiliation{School of Physics and Astronomy, University of Minnesota}
\date{\today}

\begin{abstract} 
Spin-helical states, which arise in quasi-one-dimensional (1D) channels with spin-orbital (SO) coupling, underpin efforts to realize topologically-protected quantum bits based on Majorana modes in semiconductor nanowires. Detecting helical states is challenging due to non-idealities present in real devices. Here we show by means of tight-binding calculations that by using ferromagnetic contacts it is possible to detect helical modes with high sensitivity even in the presence of realistic device effects, such as quantum interference. This is possible because of the spin-selective transmission properties of helical modes. In addition, we show that spin-polarized contacts provide a unique path to investigate the spin texture and spin-momentum locking properties of helical states. Our results are of interest not only for the ongoing development of Majorana qubits, but also as for realizing possible spin-based quantum devices, such as quantum spin modulators and interconnects based on spin-helical channels.
\end{abstract}
\maketitle


Semiconductor nanowire-superconductor hybrid systems are a highly promising materials platform for investigating Majorana zero modes (MZMs)\cite{lutchyn2010majorana, oreg2010helical, mourik2012signatures, sau2012experimental, rokhinson2012fractional,das2012zero,deng2012anomalous, albrecht2016exponential}, as well as spin transport in quantum channels \cite{yang2020spin,sun2020spin}. Because MZMs arise in odd-parity topological superconductors, realizing them in hybrid semiconductor nanowire-superconductor devices requires removing the spin degeneracy to achieve an effective odd-parity state. This can be achieved by opening a spin-helical gap in the semiconductor channel, thereby introducing a correlation between spin and momentum in certain energy ranges of the nanowire subbands \cite{lutchyn2010majorana,oreg2010helical,stvreda2003antisymmetric,pershin2004effect,quay2010observation}. A large helical gap requires strong spin-orbit (SO) materials with long mean free paths, typically InSb or InAs nanowires, in conjunction with a Zeeman interaction, usually achieved by using an external magnetic field. Despite the rapid pace of research on MZMs in nanowire-superconductor devices in recent years, there remain important open questions about the individual components. For example, a clear detection of the spin-helical state, the necessary precursor to Majorana modes in nanowire systems, is currently lacking. Conventional normal metal leads can in principle be used to investigate quantized conductance patterns that should serve as signatures of a helical gap, however imperfect quantization and quantum interference effects present in real devices lead to complex conductance patterns that cannot be unambiguously ascribed to a helical gap  \cite{kammhuber2017conductance,heedt2017signatures,rainis2014conductance,estrada2018split}. Normal metal contacts are also not suitable for studying the rich spin texture that makes the helical state relevant for realizing MZMs and for investigating other quantum spin-based effects. Recently, there has been growing interest in integrating ferromagnetic (FM) elements with high SO coupling materials to realize topological states \cite{desjardins2019synthetic,takiguchi2019giant,vaitiekenas2021zero} and to demonstrate quantum transport effects with spin-polarized currents, including spin-filtering in the regime of few quantum modes and possible spin-dependent signatures of the helical state \cite{yang2020spin,sun2020spin}. However, an approach for reliably detecting spin-helical states and exploring their spin texture is currently lacking. In this work, we model quasi-1D quantum devices semiconductor devices with strong SO and FM contacts using numerical tight-binding calculations. We find that using FM contacts can improve substantially the detectability of the helical state and circumvent limitations that arise naturally from quantum interference effects in real devices with normal contacts. Moreover, this approach can also reveal information about spin states in quantum wires, which is inaccessible with normal contacts. These results can help guide the development and interpretation of experiments on MZMs and quantum spintronics devices based on semiconductors with strong SO coupling, and could help enable future on-chip quantum interconnects for modular quantum computation \cite{debray2009all,koo2009control,chuang2015all,awschalom2021development}.

\begin{figure}[]
    \includegraphics[width=\linewidth]{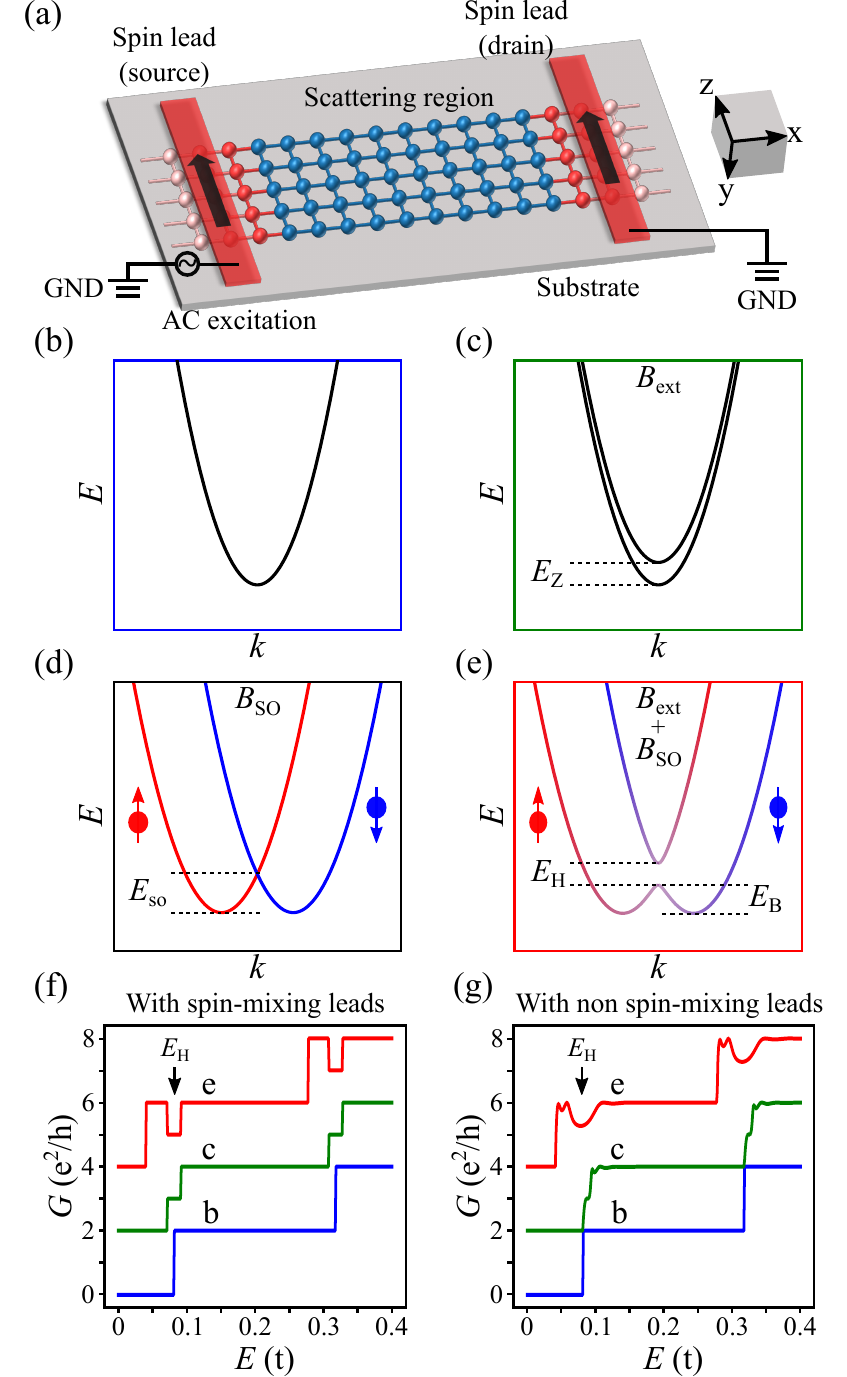}
    \caption{(a) Device structure showing the semiconductor (blue lattice) and magnetic leads (red lattice). Substrate and metallic leads are represented by the green and gray blocks respectively. (b-e) The lowest subbands of the 1D strip with (b) no SO coupling or Zeeman splitting ($\alpha=0$ and $\Bx=0$), (c) with Zeeman splitting from a finite $B_x$ ($g\mu_B\Bx=0.02t$), (d) with SO coupling (effective $\Bso$ along -z direction), and (e) with both finite Zeeman splitting and SO coupling, showing helical gap. (f-g) Conductance ($G$) vs. Fermi energy ($E$) for the three subband structures from (b), (c) and (e) respectively. (f) $G$ vs. $E$ considering normal leads with spin-mixing. (g) $G$ vs. $E$ in the absence of spin-mixing in the leads. Traces in (f) and (g) are offset vertically by $2\Go$ for clarity.}
    \label{fig1}
\end{figure}

To investigate spin-dependent transport in the helical regime, we set up a quasi-1D nanostrip with magnetic contacts on a square lattice (Fig. 1(a)) and compute the transmission matrix using KWANT \cite{groth2014kwant}.
The Hamiltonian is described by:
\begin{equation}
    \label{eq:Ham_R}
    H=\frac{p^2}{2m^\ast}-i\alpha_{\rm R}\sigma_{\rm z}\frac{\partial}{\partial x}-\frac{g\mu_{\rm B}}{2}{\vec{\sigma}}\cdot{\vec{B}},
\end{equation}
where the second term describes the SO coupling, with $\alpha_{\rm R}$ the Rashba spin-orbit coefficient and ${\sigma_{\rm z}}$ the $z$-component of the Pauli matrix, and the third term describes the Zeeman coupling to an applied magnetic field $\vec{B}$. The Rashba spin-orbit effective field $\Bso=\alpha_{\rm R}\vec{k}\times\vec{E}$ points along the $z$-direction, since $\vec{k}$ is  along $+x$, the electron propagation direction, and the electric field $\vec{E}$ is along $-y$, normal to the substrate. In our simulation, we implement the scattering region [blue lattice in Fig. 1(a)] as a $50\times10$ lattice mesh with lattice constant $a=10$ nm. It is noteworthy that although the scattering region consists of a two-dimensional lattice, the system is still effectively 1D because the electronic states are quantized into discrete levels along the strip width ($z$-direction). The resulting subbands are parabolic and behave as in a 1D conductor (see also Supplementary Section I \cite{Supp}). With this model, the hopping energy $t=\frac{\hbar^2}{2m_e^\ast a^2}=27.2$meV, where $m_e^\ast=0.014 m_e$ is the effective mass of electrons in InSb, a typical semiconductor candidate for hybrid Majorana devices. This corresponds to a nanostrip diameter of approximately 100 nm and a lead separation of approximately 500 nm, typical dimensions for experiments on InSb and InAs nanowires. To open a helical gap in the scattering region, we consider a magnetic field $B_x$ applied along the $x$-direction, with resulting Zeeman splitting $\Delta E_Z=g\mu B_x=0.02t$, and a Rashba coefficient $\alpha_R=0.4ta$. This corresponds to an applied field $B\approx200$ mT and a spin-orbit energy $E_{so}\approx1$ meV, consistent with experimental values \cite{mourik2012signatures,van2015spin}. Fig.1(b)-(e) show the nanostrip subbands computed for different combinations of magnetic fields and Rashba SO coupling. We focus here on the first subbands for clarity, but additional subbands are plotted in Supplemental Section I \cite{Supp}. The key effects remain the same as more subbands are considered. 

With the Hamiltonian set up in the lattice, we can calculate the transmission matrix of the system. We compute the transmission rate from the left lead to the right lead, which yields the conductance, $G$ after multiplying with a factor of $\Go$. In the absence of both external field and Rashba coupling, the subbands are parabolic (Fig. 1(b)) and the conductance increase step-wisely in $2e^2/h$ increments (blue curve in Fig. 1(f)). When a small finite field is applied, the subband degeneracy is lifted by the Zeeman interaction leading to a splitting of size $\Ez$ (Fig. 1(c)) and an intermediate $e^2/h$ conductance plateau develops as a result (green curve in Fig. 1(f)). In contrast, including Rashba SO coupling introduces the effective magnetic field $\Bso$ along the -$z$ direction, which shifts the subbands horizontally leading to a $k=0$ crossing at the spin-orbital energy, $\Eso$ (Fig. 1(d)). In the presence of both finite $\Bso$ and $\Bx$ a helical gap of size $\Eh$ is opened, with a pre-helical gap bump feature of size $\Eb$ (Fig. 1(e)). This results in the re-entrant conductance pattern $0 \rightarrow 2e^2/h \rightarrow e^2/h \rightarrow 2e^2/h$ (red curve in Fig. 1(f)), providing a theoretical fingerprint of the helical state in measurements with normal contacts.

The conductance traces shown in Fig. 1(f) were obtained by considering the same Hamiltonian in the leads as in the scattering region. However, this typical treatment is not suitable for the case of FM leads. We model each FM contact as two independent leads, each of which can only emit or accept one spins species, either up or down \cite{mireles2001ballistic}. The spin polarization, $P$, of the source and drain is assumed to take on the bulk value for the ferromagnet. Within the KWANT framework, FMs can be implemented by attaching one pair of leads at each end of the scattering region, which we label L$\uparrow$, L$\downarrow$ , R$\uparrow$, R$\downarrow$, for the up-spin ($\uparrow$) and down-spin ($\downarrow$) left (L) (source) and right (R) (drain) leads, respectively. This allows computation of the conductances $G_{ss'}$ between spin species $s'$ and $s$ across the device, where $s'$ and $s$ can be either $\uparrow$ or $\downarrow$. 
For the case of spin-unpolarized FM leads, which corresponds to both spin components $s$ and $s’$ having equal weights, the conductance traces are shown in Fig. 1(g). Interestingly, a softening of the conductance plateaus occurs due to conductance oscillations superimposed on the plateaus. These oscillations arise from quantum interference of partially-reflected electrons as a result of the energy mismatch between the semiconducting scattering region, in which spin-mixing is allowed, and the FM leads, in which spin-mixing is not allowed under the independent-spins FM contacts model \cite{mireles2001ballistic}. Importantly, these soft plateaus are reproduced within the typical treatment of normal leads if the spin-mixing Zeeman term vanishes in the leads, which serves as a useful test case validating our model for FM leads (more details of the implementation of the FM leads are available in Supplemental Section II \cite{Supp}). We emphasize that considering different spin-mixing properties for leads and semiconductor is not only relevant for FM leads, but also enables a more realistic description of experiments with normal leads. For example, the leads of a typical MZM device are made of a metal whose Land\'e $g$-factor (of order 2) is substantially smaller than that of the typical semiconductors (InAs or InSb, with $g$-factors of order 10-50) and thus the Zeeman splitting for a given applied field is substantially smaller in the leads. We thus expect that the red trace in Fig. 1(g) provides a more accurate depiction of the helical gap fingerprint that can be expected in the best devices with normal contacts than the commonly-considered idealized case of Fig. 1(f).

\begin{figure}
    \includegraphics[width=\linewidth]{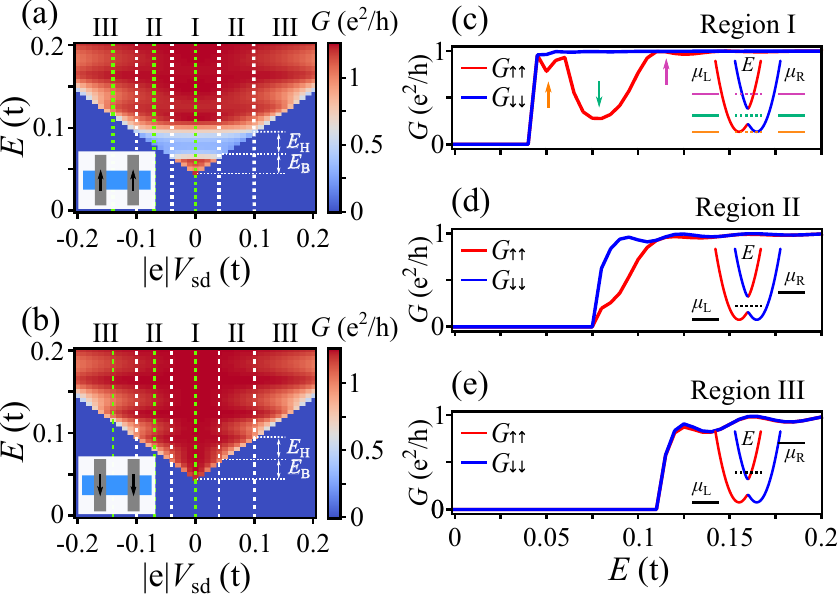}
    \caption{(a, b) Conductance of the spin-conserving channels as a function of bias window $-e\Vsd$ and Fermi energy $\E$. (a) Spin-down to spin-down conductance, $\Gdd$, (b) spin-up to spin-up conductance, $\Guu$. White dashed lines indicate the boundaries between different regions as defined in the main text. Green dashed lines denote the linecuts displayed in (c)-(e). The corresponding energy ranges of the pre-helical bump ($\Eb$) and the helical gap ($\Eh$) are labelled. (c)-(e) Vertical linecuts corresponding to $\Vsd$ values of 0, $-0.07t$, and $-0.14t$, showing (c) minor-spin conductance suppression effects (small bias), (d) spin splitting effects (intermediate bias) and (e) spin-conductance overlap (large bias) respectively. Insets in (c)-(e): subband diagrams showing the relative position of the electrochemical potentials in the leads (solid lines) and $E$ in the semiconductor scattering region (dotted lines).}
    \label{fig2}
\end{figure}

Recent experimental work \cite{yang2020spin,sun2020spin} has demonstrated that spin-filtering can be achieved in strong SO nanowire devices with magnetic contacts, however a comprehensive understanding and description of the spin transport picture in such devices is still missing. Below, we show that FM contacts can provide access to key information about the helical state and its spin texture, which cannot be obtained in devices with normal contacts. We thus expect that these insights will be of high importance for guiding future experiments. For simplicity, we begin by considering FM leads with 100\% spin polarization ($P=1$) and subsequently discuss the case of intermediate values of $P$.
We begin by examining the role of the source-drain bias voltage ($\Vsd$) on transport across the device. $\Vsd$ effects on conductance are important because spin is coupled to the net electron propagation direction in the helical gap, i.e. the group velocity. With helical spin-momentum coupling, changes in bias magnitude and sign are expected to influence spin-filtering effects, similar to the case of topological insulator devices \cite{li2014electrical,li2016direct,tian2017observation}. In the presence of an applied bias, the voltages on the two leads are given by: $\abs{e}\Vl=\E+\abs{e}\Vsd/2$ and $\abs{e}\Vr=\E-\abs{e}\Vsd/2$, where $\Mul=\abs{e}\Vl$ and $\Mur=\abs{e}\Vr$ are the electrochemical potentials of the left and right leads respectively.
We first investigate the spin-filtering effects on the spin-conserving conductance channels $\Guu$, and $\Gdd$, which can be experimentally measured by aligning the magnetizations in the two leads parallel to each other. Fig. 2(a) and Fig. 2(b) display the corresponding conductance maps as a function of $\Vsd$  and $\E$. Individual conductance traces for different values of $\Vsd$ are plotted in Fig. 2(c-e). At zero bias (Fig. 2(c)) the conductance in the $\Gdd$ channel displays a stepwise increase by $\Go$ as $E$ reaches the bottom of the subbands, then remains relatively constant with increasing energy. In contrast, the $\Guu$ channel displays a strong dip to $\sim0.3 e^2/h$. 

As a result of the helical gap, the nanowire acts as a spin filter, owing to the spin-imbalance between right-going modes (group velocity $v_{\rm g}=dE/dk>0$), which are spin-$\downarrow$, and the left-going modes ($v_{\rm g}=dE/dk<0$), which are spin-$\uparrow$ (Fig. 1(e)). Two energy scales are important to clearly illustrate the spin-filtering transport: $\Eb$, the pre-helical bump size, and $\Eh$, the helical gap size. This gives rise to three distinct regions based on the value of $\Vsd$: $|e\Vsd|/2 < \Eb$ (Region I), $\Eb<|e\Vsd|/2<\Eh+\Eb$ (Region II), and $|e\Vsd|/2>\Eh+\Eb$ (Region III). The spin-filtering conductance is readily apparent when comparing $\Guu$ with $\Gdd$ (Fig. 2). In Region I, the bias window is small enough to resolve both the pre-helical bump feature and the helical gap as $E$ is varied. In this region $\Guu$ displays a non-monotonic behavior because the right-going spin-$\uparrow$ mode is not transmitted when $E$ lies in the helical gap [green arrow in Fig. 2(c)]. When $\Vsd$ is increased further and Region II becomes accessible, the pre-helical bump feature in $\Guu$ [orange arrow in Fig. 2(c)] drops significantly below $\Go$. This occurs because once $\Mul$ reaches the subband edge from below, $E$ lies in the helical gap, where the right-going spin-$\uparrow$ modes cannot propagate. However, instead of the $\Go \rightarrow 0.3\Go \rightarrow \Go$ conductance evolution seen in $\Guu$ in Region I, Region II shows a prominent difference between $\Guu$ and $\Gdd$ as the conductance rises above 0, indicative of spin-filtering. When the bias window is increased further (Region III), for similar reasons, once $\Mul$ reaches the bottom of the subbands (Fig. 2(e) inset), $E$ crosses above the helical gap and consequently the $\Guu$ trace has a stepwise $\Go$ increment. It is useful to compare the above results for FM leads to the conventional normal-contact conductance measurement technique. Whereas with normal contacts it is necessary to consecutively resolve both the pre-helical-bump ($2\Go$ conductance) and the helical gap (close to $\Go$ conductance) with a small bias window in Region I, the use of FM contacts leverages the spin-filtering properties of the helical gap to enable its clear detection over a large range of $\Vsd$ extending over both Region I and Region II.

\begin{figure}
    \includegraphics[width=\linewidth]{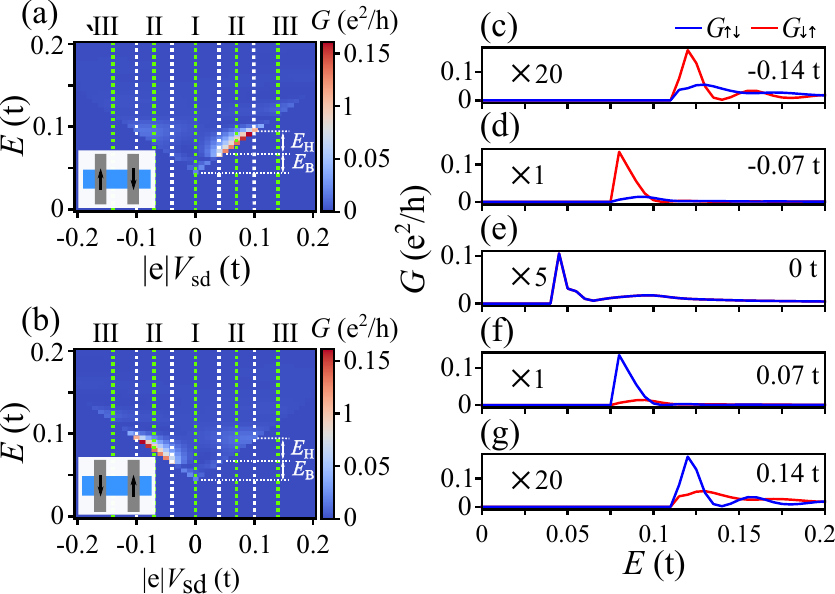}
    \caption{(a, b) Conductance of the spin-flip channels as a function of bias window $-e\Vsd$ and energy. (a) Spin-up to spin-down ($\Gdu$). (b) Spin-down to spin-up ($\Gud$). White dashed lines mark the boundaries of the regions defined as in Figure 2. Green dash lines denote the axis along which (c-g) are extracted. The corresponding energy ranges of the pre-helical bump ($\Eb$) and the helical gap ($\Eh$) are labelled. (c-g) Vertical linecuts for $-e\Vsd$ values of $-0.14t$, $-0.07t$, 0, $0.07t$, and $0.14t$, highlighting the evolution of the bias-asymmetric transport feature in the spin-flip transport channels.}
    \label{fig3}
\end{figure}

When $E$ is located in the  the helical gap, the spin is coupled to the electron propagation direction, which suggests that the spin-filtering properties of the system should depend on the sign of $\Vsd$. Fig. 2(a,b) show that the spin-conserving conductance ($\Guu$, $\Gdd$) is independent of bias polarity. To observe a dependence on bias polarity it is necessary to consider the two spin-flip channels, $\Gdu$ and $\Gud$, which describe the conductance from $L\uparrow$ to $R\downarrow$ and $L\downarrow$ to $R\uparrow$ respectively. For fully spin-polarized leads, this corresponds to experimental conductance measurements with antiparallel magnetizations in the two FM leads. Surprisingly, although the bias-asymmetric effects are missing in the spin-conserving transport channel, a substantial correlation between bias polarity and spin orientation exists in the spin-flip channel.  Fig. 3(a,b) show the corresponding conductance maps for the lowest subband as a function of $\E$ and $\Vsd$, revealing a clear bias-polarity dependence. For example, Fig. 3(a) shows the evolution of $\Gdu$ ($L\uparrow$ to $R\downarrow$), revealing an enhancement of the spin-flip conductance along a ridge at the edge of Region II, where the lower electrochemical potential is aligned to the subband bottom and and $E$ is in the helical gap. An intuitive understanding of the bias polarity dependence is that the bias window applied on the two leads drives electrons moving from the higher electrochemical potential to the lower one, resulting in an effective $\vec{k}$ along this direction. This consequently influences the $\Bso$ direction: for example, in Fig. 3(a), the positive bias window leads to an effective $\vec{k}$ along $+x$ ($\Bso$ along $-z$), thus increasing the spin-up to spin-down transmission. Similar effects  occur for $\Gud$ under reversed bias polarity. The exact mechanisms leading to the observed ridge resonances and their bias asymmetry require further study, however this numerical observation suggests a novel experimental signature of the helical state that arises only in devices with FM contacts.

\begin{figure}
    \includegraphics[width=\linewidth]{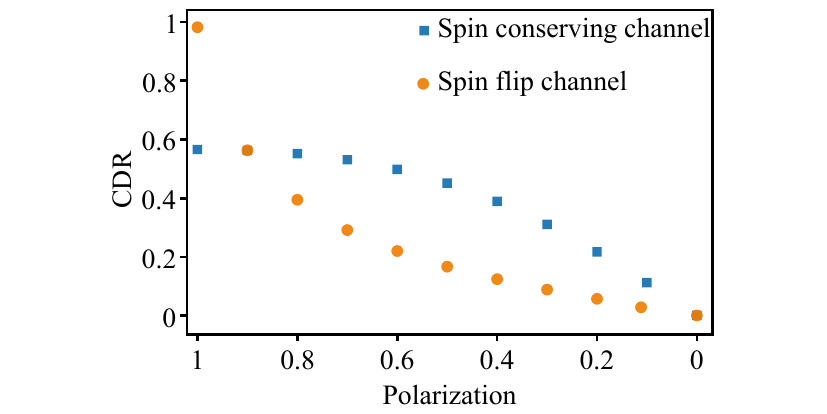}
    \caption{Dependence of the conductance difference ration (CDR) on spin polarization, for the spin-conserving channel (blue squares) and the spin-flip channel (orange disks).}
    \label{fig4}
\end{figure}

In the preceding discussion, we have considered fully spin-polarized leads. We now evaluate the impact of finite $P$ on the spin conductance contrast. To model spin-polarized transport we introduce the constants $\tmaj$ ($\tmin$), with $\tmaj>\tmin$ and $\tmaj+\tmin=1$, which describe the coupling of the majority (minority) spins from the leads to the semiconductor region. Majority (minority) spins are aligned parallel (anti-parallel) to the lead’s magnetization. The polarization of the leads is defined as $P=\frac{\tmaj-\tmin}{\tmaj+\tmin}$ and describes the imbalance between the two spin channels in transport. A normal (non-FM) lead has $P=0$ and $\tmaj=\tmin=0.5$; a fully spin-polarized lead has $P=1$, $\tmaj=1$ and $\tmin =0$. Within this framework we calculate the spin-dependent conductance by using the coupling constants as weights for each of the two spin-conserving and each of the two spin-flip channels (calculation details and conductance color plots are shown in the supplementary section III \cite{Supp}). For simplicity, we consider the typical case where the left and right leads have the same spin polarization, but this assumption can be relaxed in a straightforward manner within our model.

In order to quantitatively evaluate the impact of finite spin polarization on the spin-resolved conductance for parallel and antiparallel lead magnetizations, we compute the conductance difference ratio (CDR), defined as the maximum of $\frac{\abs{G_1-G_2}}{G_1+G_2}$ extracted from the conductance maps as shown in Fig. 2(a,b) and Fig. 3(a,b) for $P=1$, where $G_1$ and $G_2$ correspond to $\Guu$ and $\Gdd$ for the spin-conserving channel (measured with parallel magnetization), and to $\Gud$ and $\Gdu$ for the spin-flip channel (measured with antiparallel magnetization). Fig. 4 displays the dependence of the CDR on $P$ for the spin-conserving and spin-flip channels. For the spin-conserving channel, the CDR has a maximum value of ${\rm CDR}=0.57$, at $P=1$, and decreases gradually with decreasing $P$ to reach $80\%$ of its maximum value at $P=0.5$, and 0 for completely unpolarized (normal) leads. In contrast, for the spin-flip channel, the CDR drops rapidly from a value of 1, at $P=1$, reaching less than $20\%$ of its maximum value at $P=0.5$. Thus, the bias-asymmetric signatures of the helical state, which as discussed earlier arise only in the spin-flip channel, depend crucially on the presence of a substantial spin polarization, although a finite contrast remains present for any non-zero value of $P$ provided FM leads are used.

\begin{figure}
    \includegraphics[width=\linewidth]{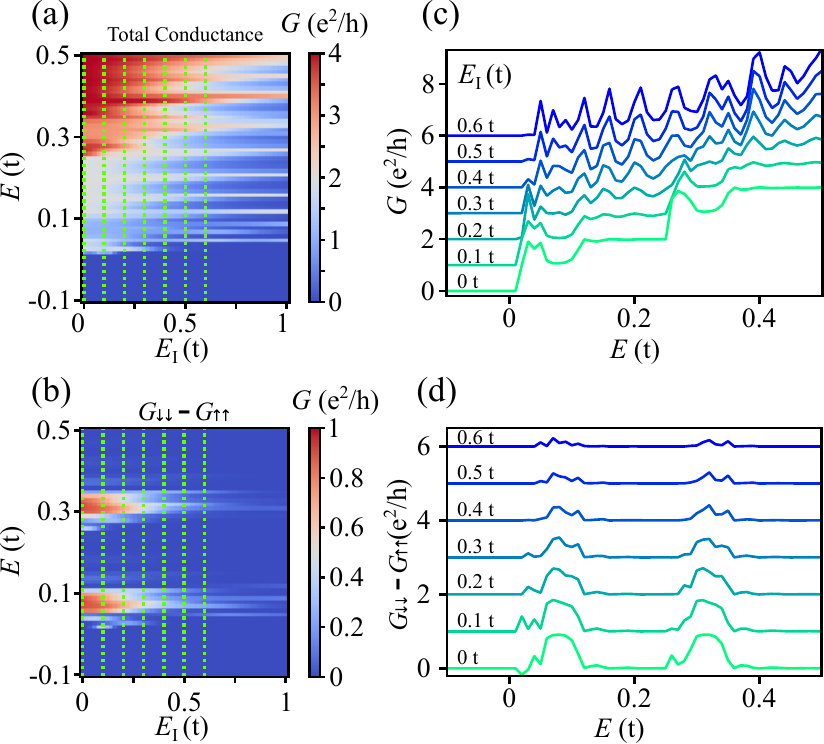}
    \caption{(a) Total conductance map for normal metallic leads. (b) Conductance difference between the major spin transport channel ($\downarrow$ to $\downarrow$) and the minor channel ($\uparrow$ to $\uparrow$) for FM leads. (c) Vertical linecuts extracted from (a) for different interfacial barrier heights. (d) Vertical linecuts extracted from (b) for different interfacial barrier heights. Linecuts are indicated by green dash lines in the color plots. The traces in (c) and (d) are offeset vertically for clarity.}
    \label{fig5}
\end{figure}

In 1D quantum systems with partially-reflecting ends the interference of propagating and reflected particle waves leads to quantum interference, analogous to that of photons in a laser cavity. Such Fabry-Pérot intereference is common for electron waves in mesoscopic systems, such as nanotube- and nanowire-based quantum devices \cite{liang2001fabry, kretinin2010nwFabry, sahoo2005electric, man2006spin, yang2020spin}, where the electrostatic potential can be spatially dependent in the neighborhood of metal/semiconductor interfaces, leading to partial reflection of the electron wavefunctions, e.g. near the contacts. Fabry-Pérot interference can severely limit or even prevent the detection of helical gaps using normal (non-FM) leads by limiting the visibility of conductance plateaus or the ability to distinguish conductance fingerprints of helical states from interference-related conductance modulation \cite{kammhuber2017conductance,heedt2017signatures,rainis2014conductance,estrada2018split,yang2020spin}. Below, we demonstrate that these challenges can be greatly mitigated by using FM contacts, which can substantially improve the contrast between helical features and quantum interference. To investigate Fabry-Pérot intereference, we now introduce a finite energy barrier at the interface between leads and nanostrip by increasing the onsite potential energy for the interfacial lattice sites (one-layer of lattice along the $z$-direction in the scattering region next to the leads) by a value $\Ei$, which we tune between $0$ to $t$. This induces partial reflection at the lead-semiconductor interface, which leads to characteristic Fabry-Pérot interference oscillations in the transmission as a function of energy, as shown in Fig. 5. The effects of the interference are very different, depending on whether the leads are normal or FM. Since helical gap detection using normal leads is based on using the sequence of conductance steps associated with the gap as the distinguishing feature (Fig. 1(f)), this method requires a clear observation of: i) the subband pre-helical bump (conductance at 2$\Go$), ii) the $\Go$ conductance suppression in the helical gap and iii) the subsequent increase back to 2$\Go$. However, as shown in Fig. 5(a), where the total conductance is computed, for normal contacts this conductance fingerprint is engulfed by quantum interference oscillations even for interfacial barriers as low as ~0.2$t$. For even stronger potential distortions near the contacts, the interference-induced oscillations make it all but impossible to resolve the conductance fingerprint of the helical gap. In contrast, by using two FM contacts in a spin-valve configuration, the spin filtering characteristics of the helical gap can be leveraged, allowing the helical gap to be detected across a substantially wider range of interfacial barrier strengths (Fig. 5(b)). This is because of the contrast between the conductances $\Guu$ and $\Gdd$, which can be obtained experimentally by comparing the conductances obtained before and after flipping the magnetization directions of both FM contacts. Thus, using FM contacts not only enables detection of the spin properties of the helical gap, but it can also substantially alleviate the difficulties in detecting the gap in the first place. These difficulties are largely unavoidable in the conventional, normal-contact devices under realistic conditions because quantum interference is nearly always present. The calculations presented here were performed for fully polarized leads, $P=1$. For $P < 1$, the signal shown in Fig. 5(b,d), which is obtained from the conductance difference in the spin-conserving channels, will decrease slowly following the tendency shown in Fig. 4, but an advantage over normal leads is still expected for $P > 0$.

In this work we modeled quantum transport in the helical gap regime of semiconductor nanowires with strong Rashba spin-orbital interaction and spin-selective contacts by using a one-dimensional tight-binding model. Our quantitative numerical results show that FM contacts can be used to successfully detect the helical gap in the presence of realistic device effects that would prohibit detection with conventional normal contacts. Moreover, we show that magnetic contacts provide access to new physics associated with the spin structure of the helical gap. In particular, we found that spin-conserving and spin-flip transport are affected in qualitatively distinct ways by applied bias, bias polarity and interface barriers. This work reveals the importance of experimentally integrating FM elements, especially materials with high spin polarization, such as certain Heusler alloys, within nanowire-based topological quantum devices for creating a new generation of robust experiments on Majorana zero-modes and for developing 1D quantum spintronic devices. 

We acknowledge B. Heischmidt and J. Etheridge for helpful discussions. This work was supported by the Department of Energy under award no. DE-SC-0019274.

\bibliography{main}
\end{document}